\documentclass[twocolumn,showpacs,preprintnumbers,amsmath,amssymb,prl]{revtex4-1}
\usepackage{epsfig} %Works faster than the graphicx package}
\usepackage{dcolumn}% Align table columns on decimal point
\usepackage{bm}% bold math

\begin{document}

\preprint{\today} % This adds a date to the top right corner of the paper

\title{Determination of the scalar and vector polarizabilities of the cesium $6s \ ^2S_{1/2} \rightarrow 7s \ ^2S_{1/2}$ transition and implications for atomic parity non-conservation}

\author{George Toh$^{1,2}$, Amy Damitz$^{2,3}$, Carol E. Tanner$^4$, W. R. Johnson$^4$, and D. S. Elliott$^{1,2,3}$}

\affiliation{%
   $^1$School of Electrical and Computer Engineering, Purdue University, West Lafayette, Indiana 47907, USA\\
   $^2$Purdue Quantum Science and Engineering Institute, Purdue University, West Lafayette, Indiana 47907, USA\\
   $^3$Department of Physics and Astronomy, Purdue University, West Lafayette, Indiana 47907, USA\\
   $^4$Department of Physics, University of Notre Dame, Notre Dame, Indiana 46556, USA
}

\date{\today}% It is always \today,

\begin{abstract}
Using recent high-precision measurements of electric dipole matrix elements of atomic cesium, we make an improved determination of the scalar ($\alpha$) and vector ($\beta$) polarizabilities of the cesium $6s \ ^2S_{1/2}  \rightarrow 7s \ ^2S_{1/2} $ transition calculated through a sum-over-states method. We report values of $\alpha = -268.82 \ (30) \ a_0^3$ and $\beta = 27.139 \ (42) \ a_0^3$ with the highest precision to date. We find a discrepancy between our value of $\beta$ and the past preferred value, resulting in a significant shift in the value of the weak charge $Q_w$ of the cesium nucleus.  Future work to resolve the differences in the polarizability will be critical for interpretation of parity non-conservation measurements in cesium, which have implications for physics beyond the Standard Model.
\end{abstract}
                             
\maketitle 

Precision measurements of weak optical interactions in atoms can provide a sensitive means of probing the weak force between nucleons and electrons at low momentum transfer~\cite{BouchiatB74a, BouchiatB74b}.  
The extent to which atomic parity non-conservation (PNC) measurements agree with standard model predictions can provide constraints on conjectures of `beyond standard model' physics, which are based on new additional interactions involving, for example, a massive $Z^{\prime}$ boson~\cite{PorsevBD09,PorsevBD10,DzubaBFR12,ErlerL00,DienerGT12}, a light boson~\cite{BoehmF04,BouchiatF05,DzubaFS17}, or axion-like particles~\cite{StadnikDF18}, or searches of dark energy~\cite{RobertsSDFLB14b, RobertsSDFLB14a, StadnikF14a, StadnikF14b}.
Recent theoretical searches for dark matter~\cite{DavoudiaslLM12, DavoudiaslL14, DavoudiaslLM14} are based on a hypothesized light dark boson that decays primarily to dark matter, but which also interacts weakly with standard model matter.  

The most precise determination of the weak charge through atomic PNC measurements to date was carried out in atomic cesium.  This determination is based on a precise measurement of the ratio $Im ( \mathcal{E}_{\rm PNC}) / \beta$ by Wood {\it et al.}~\cite{WoodBCMRTW97}, where $Im ( \mathcal{E}_{\rm PNC})$ is the electric dipole transition moment for the $6s \ ^2S_{1/2} \rightarrow 7s \ ^2S_{1/2}$ transition induced by the weak force interaction, and $\beta$ is the vector polarizability for the transition.  The weak charge $Q_w$ is determined then as the product of $Im ( \mathcal{E}_{\rm PNC}) / \beta$, the polarizability $\beta$, and a proportionality factor $k_{PNC} \equiv Q_w / Im ( \mathcal{E}_{\rm PNC})$, which must be determined through difficult atomic structure calculations~\cite{DzubaFS89, BlundellJS91, BlundellSJ92, Derevianko00, DzubaFG01, JohnsonBS01, KozlovPT01, DzubaFG02, FlambaumG05, PorsevBD09, PorsevBD10, DzubaBFR12, RobertsDF2013}.
A new determination of $Im ( \mathcal{E}_{\rm PNC}) / \beta$ is currently under development in our laboratory, and Derevianko has announced plans to undertake a new calculation of $k_{PNC}$~\cite{WiemanD19}.  In this paper, we report a new determination of the vector polarizability $\beta$, which is of higher precision than, but differs from, the previously accepted value~\cite{DzubaF00, BennettW99}.

Since 2000, the most precise determination of $\beta$ has been based upon a theoretical value for the hyperfine-changing magnetic dipole matrix element $M1_{\rm hf}$~\cite{DzubaF00}, and a laboratory determination of the ratio $M1_{\rm hf} / \beta$ \cite{BennettW99}.  
With a precision of 0.19\%, this value of $\beta$ has been preferred over the value determined from a calculation of the scalar polarizability $\alpha$ using a sum-over-states approach~\cite{BlundellSJ92,SafronovaJD99, VasilyevSSB02, DzubaFG02}, combined with a measurement of the ratio $\alpha / \beta$ \cite{ChoWBRW1997}.  The latter method requires precise measurements or theoretical values for the reduced electric dipole (E1) matrix elements $\langle np_J || r || ms_{1/2}\rangle$ with $m = 6$ or $7$, $n \ge 6$ and $J = 1/2$ or $3/2$.  Many of these matrix elements were measured to great precision in the past thirty years~\cite{BouchiatGP84, TannerLRSKBYK92, YoungHSPTWL94, RafacT98, RafacTLB99, BennettRW99, VasilyevSSB02, DereviankoP02a, AminiG03, BouloufaCD07, SellPEBSK11, ZhangMWWXJ13, antypas7p2013, Borvak14, PattersonSEGBSK15, GregoireHHTC15}, and in the last 3 years, our group has undertaken and completed high-precion measurements of the remainder of these eight matrix elements ~\cite{TohJGQSCWE18, TohDGQSCSE19, DamitzTPTE18a}.

We first present a new determination of $\alpha$ through a sum-over-states method~\cite{BlundellSJ92, VasilyevSSB02}
\begin{eqnarray}
  \alpha &=& \frac{1}{6} \sum_{n} \Bigg[ \langle 7s_{1/2} || r || np_{1/2} \rangle \langle np_{1/2} || r || 6s_{1/2} \rangle \nonumber \\
  && \hspace{0.2in} \times \left( \frac{1}{E_{7s} - E_{np_{1/2}}} + \frac{1}{E_{6s} - E_{np_{1/2}} }\right) \nonumber \\
  && - \langle 7s_{1/2} || r || np_{3/2} \rangle \langle np_{3/2} || r || 6s_{1/2} \rangle  \nonumber \\
   && \hspace{0.2in} \times \left( \frac{1}{E_{7s} - E_{np_{3/2}}} + \frac{1}{ E_{6s} - E_{np_{3/2}} }\right) \Bigg]
   \label{eqn:alphasum}
\end{eqnarray}
where $\langle np_J || r || ms_{1/2}\rangle$ are the E1 transition matrix elements, $E_{ms}$ and $E_{np_{J}}$ are state energies, and $J = 1/2$ or $3/2$ is the electronic angular momentum of the state. 

\begin{table*}[t!]
  \caption{E1 dipole matrix elements, eigenstate energies, and contributions to the scalar polarizability $\alpha$. This table shows our sum-over-states calculation, as given in Eq.~(\ref{eqn:alphasum}), of the scalar polarizability $\alpha$. E1 elements for $n = 6$ and $7$, 
  %above the dashed line, 
  are experimental values, as discussed in the text. $^a$Refs.~\cite{TohJGQSCWE18, TohDGQSCSE19}, $^b$Refs.~\cite{YoungHSPTWL94,RafacT98,RafacTLB99, DereviankoP02a, AminiG03,BouloufaCD07,ZhangMWWXJ13, PattersonSEGBSK15,GregoireHHTC15, TannerLRSKBYK92, SellPEBSK11}, $^c$Ref.~\cite{BennettRW99} and this work, $^d$Ref.~\cite{DamitzTPTE18a}. Theory values of E1 elements ($8 \le n \le 12$) are from Ref.~\cite{SafronovaSC16} including the Supplemental Information.  State energies are found in NIST tables~\cite{kramida2016nist}.}
  \label{table:alphasumoverstates}
  \begin{tabular}{c|ccc|ccc|c|rr}
    \hline
$n$ &	$d \ (a_0)$	&   $\delta d$(\%)	&  $\delta \alpha \ (a_0^3)$	 
    &   $d \ (a_0)$	&   $\delta d$(\%)	&  $\delta \alpha \ (a_0^3)$   
    &	$E_{np_{1/2}}$ (cm$^{-1})$	& \multicolumn{1}{c}{$\alpha \ (a_0^3)$}  &	\multicolumn{1}{c}{$\delta \alpha \ (a_0^3)$}  \\ \hline

\rule{0in}{0.15in} & \multicolumn{3}{c|} {$\langle 7s_{1/2} || r || np_{1/2} \rangle$} & \multicolumn{3}{c|} {$\langle np_{1/2} || r || 6s_{1/2} \rangle$} & & \\

\hspace{0.05in}$6$\hspace{0.05in} &	\hspace{0.05in}$-4.249^a$ \hspace{0.05in} & \hspace{0.05in}$0.094$\hspace{0.05in}       &	\hspace{0.05in}$0.031$\hspace{0.05in}	
    &   \hspace{0.05in}$4.5057^b$\hspace{0.05in}  &  \hspace{0.05in}$0.035$\hspace{0.05in}	  & \hspace{0.05in}$0.011$\hspace{0.05in}	
    & \hspace{0.05in}$11178.27$\hspace{0.05in} &  \hspace{0.05in} $-32.54$      &	 \hspace{0.05in}$0.03$   \\
    
$7$ &  $10.325^c$   & $0.05$	&  $0.019$	
	&  $0.2781^d$   & $0.16$	&  $0.060$
    &  $21765.35$ & $-37.35$	&  $0.06$    \\   %\cdashline{1-10}

$8$ &  $0.914$	  &  $2.9$      &  $0.016$  
	&  $0.092$	  &  $11$	    &  $0.061$
    &  $25708.84$ &  $-0.55$    &  $0.06$  \\ 
    
$9$ &  $0.349$	&  $2.9$		&  $0.002$
	&  $0.043$	&  $16$			&  $0.013$
    &  $27637.00$ & $-0.08$		&  $0.01$  \\
    
$10$ & $0.191$  &  $3.1$		&  $0.001$ 
	&  $0.025$  &  $20$ 		&  $0.005$
    &  $28726.81$  &  $-0.02$	&  $0.00$   \\
    
$11$  & $0.125$ &  $3.5$		&  $0.000$ 
	&  $0.016$  &  $27$ 		&  $0.002$
    &  $29403.42$  &  $-0.01$   &  $0.00$   \\

$12$  & $0.09$ 		&  $3.9$	&  $0.000$ 
	&  $0.012$  	&  $28$ 	&  $0.001$
    &  $29852.68$  	&  $-0.00$  &  $0.00$   \\
    
\rule{0in}{0.15in} & \multicolumn{3}{c|} {$\langle 7s_{1/2} || r || np_{3/2} \rangle$} & \multicolumn{3}{c|} {$\langle np_{3/2} || r || 6s_{1/2} \rangle$} & & \\
    
$6$	&  $-6.489^a$  	&  $0.077$ 	 &  $0.072$
	&  $-6.3398^b$   	&  $0.035$   &  $0.033$  
    &  $11732.31$  	&  $-92.93$  &  $0.08$   \\
    
$7$ &  $14.344^c$    &  $0.05$  	&  $0.051$  
	&  $-0.5742^d$   &  $0.10$	&  $0.101$
    &  $21946.39$  &  $-102.05$	&  $0.11$  \\   %\cdashline{1-10} 
    
$8$ &  $1.62$		&  $2.2$	&  $0.053$
	&  $-0.232$		&  $6.2$	&  $0.151$
    &  $25791.51$  	&  $-2.43$	&  $0.16$ \\

$9$ &  $0.68$  		& $2.1$		&  $0.010$  
	&  $-0.130$ 	& $7.4$		&  $0.035$	
    &  $27681.68$  	& $-0.47$   &  $0.04$  \\
    
$10$ &  $0.396$  	&  $2.2$	&  $0.004$  
	&  $-0.086$  	&  $8.3$	&  $0.014$
    &  $28753.68$  	&  $-0.17$  &  $0.01$  \\
    
$11$ &  $0.270$  	&  $2.4$	&  $0.002$
	&  $-0.063$   	&  $8.9$	&  $0.007$
    &  $29420.82$	&  $-0.08$ 	&  $0.01$ \\	
    
$12$ &  $0.201$  	&  $3.7$	&  $0.002$
	&  $-0.049$   	&  $9.5$	&  $0.004$
    &  $29864.54$	&  $-0.04$ 	&  $0.00$ \\	    
    
\multicolumn{7}{c}{} & \multicolumn{1}{r}{$\alpha_{n > 12} = $}  &  \rule{0in}{0.15in} $-0.30$  &  $0.15$ \\
\multicolumn{7}{c}{} & \multicolumn{1}{r}{$\alpha_{vc} = $}  &   \underline{$+0.2$}  &  $0.1$ \\
\multicolumn{7}{c}{} & \multicolumn{1}{r}{$\alpha = $}  &   $-268.82$  &  $0.30$ \\
%    \hline
  \end{tabular}
\end{table*}

We show the E1 matrix elements $\langle 7s_{1/2} || r || np_J \rangle$ and $ \langle np_J || r || 6s_{1/2} \rangle$, and state energies $E_{np_{J}}$ for states with principal quantum number $6 \le n \le 12$ used for our sum-over-states calculation in Table \ref{table:alphasumoverstates}. 
In earlier calculations of $\alpha$ \cite{SafronovaJD99, VasilyevSSB02}, the terms contributing the most to the $0.4\%$ uncertainty in $\alpha = 269.7 (11)~a_0^3$ were the $\langle 7s_{1/2} || r || 6p_{J} \rangle$ and $\langle 7p_{J} || r || 6s_{1/2} \rangle$ matrix elements whose uncertainties at that time were $0.5 \%$ and $0.6 \%$, respectively. (The numbers in brackets following the value denote the 1~$\sigma$ uncertainty in the least significant digits.)
In the following paragraphs, we summarize the recent contributions towards these matrix elements, which enable us to calculate a more precise value for $\alpha$.

\paragraph{6s-6p} The values for the $\langle 6s_{1/2} || r || 6p_{J} \rangle$ matrix elements have been measured precisely in a variety of experiments. These include fast-beam laser~\cite{TannerLRSKBYK92,RafacTLB99}, time-resolved fluorescence~\cite{YoungHSPTWL94}, ultra-fast pump-probe laser~\cite{PattersonSEGBSK15}, photoassociation~\cite{DereviankoP02a, BouloufaCD07, ZhangMWWXJ13}, ground-state polarizability~\cite{AminiG03} and atom interferometry~\cite{GregoireHHTC15}. We take the weighted average of these measurements, to obtain a precision of $\sim 0.035 \%$ for these matrix elements.

\paragraph{7s-6p} In 2017, we used an asynchronous gated detection technique with a single-photon detector to measure the lifetime of the $7s$ state to a precision of $0.14\%$ \cite{TohJGQSCWE18}. We combine this high precision lifetime measurement with a measurement of the ratio of dipole matrix elements $\langle 7s_{1/2} || r || 6p_{3/2} \rangle/\langle 7s_{1/2} || r || 6p_{1/2} \rangle$ \cite{TohDGQSCSE19} in order to determine the individual matrix elements to a precision of $<0.1\%$. This ratio measurement was based upon measurements of the influence of laser polarization on the two-photon $6s \rightarrow 7s$ transition rate.

\paragraph{7s-7p} We derive new values for the $7s-7p$ matrix elements from a dc Stark shift $\Delta \alpha_{6s7s}$ measurement of the $6s \rightarrow 7s$ transition \cite{BennettRW99}, and our high precision determinations of the $7s-6p$ matrix elements. This is the same method as used in Ref.~\cite{SafronovaJD99}.  The static polarizability $\alpha_{7s}$ depends primarily on the $7s-7p$ and $7s-6p$ values. We use $\Delta \alpha_{6s7s}$ \cite{BennettRW99} and high precision measurements of the ground state static polarizability $\alpha_{6s}$ \cite{AminiG03, GregoireHHTC15} to calculate the static polarizability $\alpha_{7s}$ of the $7s$ state. We also use theoretical calculations of the ratio of $7s-7p_J$ matrix elements $R_{7s7p} = |\langle 7s_{1/2} || r || 7p_{3/2} \rangle/\langle 7s_{1/2} || r || 7p_{1/2} \rangle| = 1.3892 \: (3)$ \cite{SafronovaJD99} and for the $7s-np$ matrix elements where $n>7$ \cite{SafronovaSC16}. The results of our determination are $\langle 7s_{1/2} || r || 7p_{1/2} \rangle = 10.325 \: (5)~a_0$ and $\langle 7s_{1/2} || r || 7p_{3/2} \rangle = 14.344 \: (7)~a_0$, an improvement in precision from $0.15\%$ in \cite{SafronovaJD99} to $0.05\%$ as presented here. 

\paragraph{6s-7p} 
Most recently, we have completed a comprehensive study of the $6s \rightarrow 7p_{3/2}$ ($\lambda = 456$ nm) and  $6s \rightarrow 7p_{1/2}$ ($\lambda = 459$ nm) line ab\-sorp\-tion strengths to determine the transition matrix elements $\langle 6s_{1/2} || r || 7p_{3/2} \rangle$ and $\langle 6s_{1/2} || r || 7p_{1/2} \rangle$ ~\cite{DamitzTPTE18a}. These comparative studies yield the ratios of matrix elements $\langle 6s_{1/2} || r || 6p_{1/2} \rangle/\langle 6s_{1/2} || r || 7p_{3/2} \rangle$ and $\langle 6s_{1/2} || r || 7p_{3/2} \rangle/\langle 6s_{1/2} || r || 7p_{1/2} \rangle$.  Then by using the very precise value of $\langle 6s_{1/2} || r || 6p_{1/2} \rangle$~\cite{TannerLRSKBYK92, RafacT98, YoungHSPTWL94, PattersonSEGBSK15, RafacTLB99, DereviankoP02a, BouloufaCD07, ZhangMWWXJ13, AminiG03, GregoireHHTC15}, we obtain a value of $\langle 6s_{1/2} || r || 7p_{3/2} \rangle$ with 0.10\% uncertainty, and of $\langle 6s_{1/2} || r || 7p_{1/2} \rangle$ with 0.16\% uncertainty.

\begin{figure}[tb]
\begin{center}
 \includegraphics[width=8.5cm]{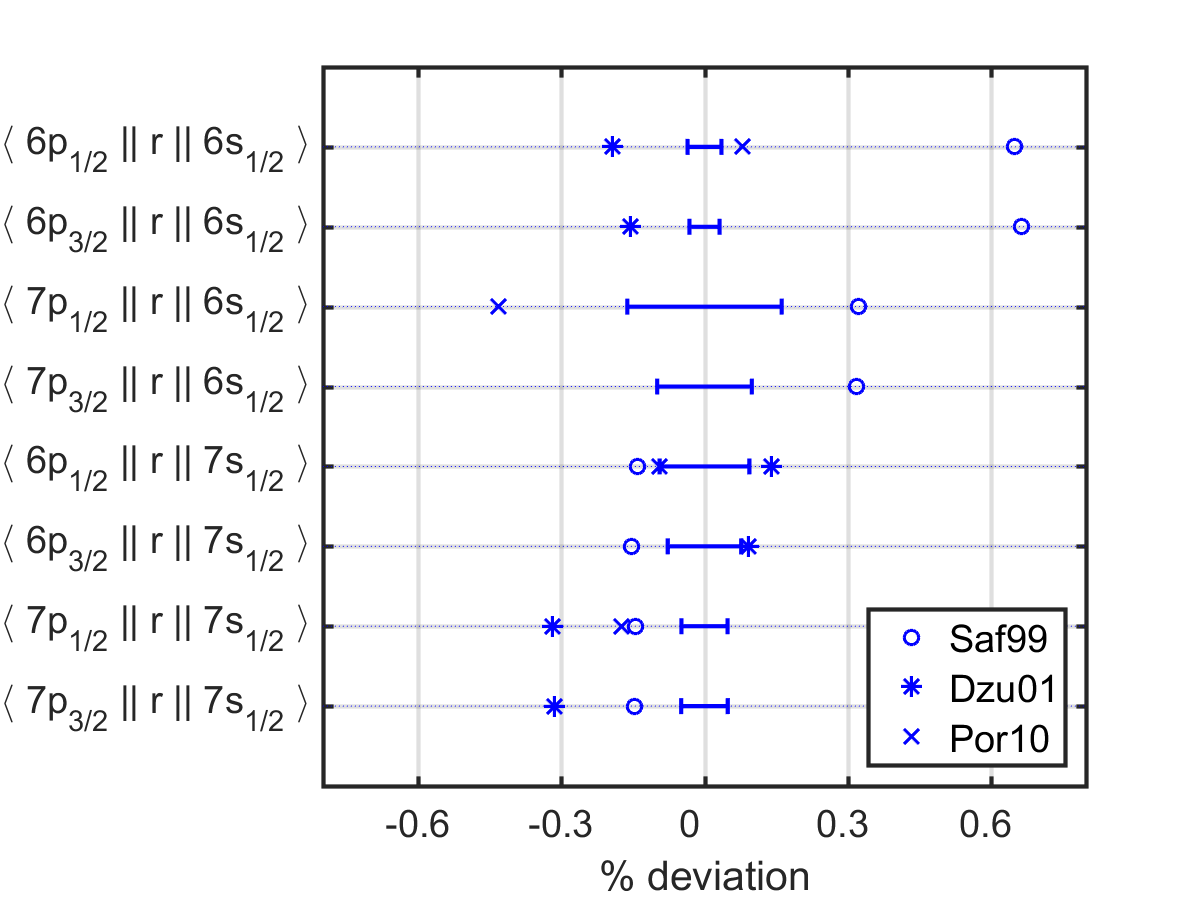}
\end{center}
    \caption{A graphical summary of the current status of the measured and calculated matrix elements $\langle np_J || r || ms_{1/2}\rangle$, where $m, n = 6$ or 7, and $J = 1/2$ and $3/2$ in atomic cesium.  The error bars show the magnitudes of the uncertainty of the measurements. The data points show the deviation between the most recent calculations of the matrix elements and the measured value. (Deviation $>0$ indicates the theoretical value is greater than the experimental value.)  The calculated values are from Refs.~\cite{SafronovaJD99} ($\circ$), \cite{DzubaFG01,DzubaFG02} ($*$), and \cite{PorsevBD10} ($\times$). }
	  \label{fig:pncmatrixelements}
\end{figure}

In Fig.~\ref{fig:pncmatrixelements} we show a plot that illustrates the current state of theory and experiment for these eight matrix elements. (This plot is an updated version of a plot that first appeared as Fig.~2 of \cite{PorsevBD09}.) Specifically, this plot shows the experimental uncertainties and the discrepancies between theory and experiment for selected transition matrix elements.  
The error bars indicate the experimental uncertainties, while markers show the difference between experiment and three recent theoretical works, including: Refs.~\cite{SafronovaJD99} ($\circ$), \cite{DzubaFG01,DzubaFG02} ($*$), and \cite{PorsevBD10} ($\times$). 
(Deviation $>0$ indicates the theoretical value is greater than the experimental value.) We observe that there is good agreement between experiment and theory to the $\sim0.2\%$ level for most of these terms.
All of the matrix elements $\langle ns_{1/2} || r || mp_{J} \rangle$ for $n, m = 6, 7$ have now been measured to a precision of 0.16\% or better, clearing the way for a new determination of $\alpha$, and serve as important benchmarks for future atomic theory calculations of $k_{PNC}$.

Table \ref{table:alphasumoverstates} shows a term-by-term computation of the scalar polarizability $\alpha$ following Eq.~(\ref{eqn:alphasum}). In the second and fifth columns, we list values of the E1 matrix elements $d = \langle 7s_{1/2} || r || np_J \rangle $ and $\langle np_J || r || 6s_{1/2}\rangle$, respectively, for principal quantum number $n$. 
    For $n = 6$ and 7, we have already discussed the values that we use.  For $n = 8-12$, we use theoretical values of these matrix elements from Ref.~\cite{SafronovaSC16}.  The signs of these matrix elements are consistent with the sign convention described in Refs.~\cite{DzubaFS97, TohDGQSCSE19}.  In each case, the percentage uncertainty of the matrix element $\delta d$ is listed in columns 3 and 6.  We show in column nine the contribution of these elements to the scalar polarizability, using the energy of $np_J$ states listed in the table~\cite{kramida2016nist}, and $E_{7s} = 18535.53$ cm$^{-1}$.  We also show the uncertainties $\delta \alpha$ resulting from $\delta d$ in this table; $\delta \alpha$ due to the uncertainty in  $\langle 7s_{1/2} || r || np_J \rangle $ in column four and $\langle np_J || r || 6s_{1/2}\rangle$ in column seven, and the quadrature sum of these in the final column.  
    
    The final contributions to $\alpha$ are from $np_J$ states with $n > 12$, and valence-core contributions $\alpha_{vc}$. We calculate the contributions from Hartree-Fock (HF) bound state wavefunctions with $n > 12$ (bound and continuum) with the aid of a B-spline basis set.  The HF value $\alpha_{n>12} = -0.45 \ a_0^3$ is obtained by subtracting the sum for $n=1$ to $12$, in a term-by term HF calculation, from the sum over the entire spline basis.  Noting that the HF values for the known contributions to $\alpha$ for $n = 6$ to $12$ are typically $\sim$30\% too high, we estimate $\alpha_{n>12} = -0.30 \ (15) \ a_0^3$. 
    For the valence-core contributions, we determine $\alpha_{vc} = +0.2 \ (1) \ a_0^3$, in agreement with the value reported in \cite{VasilyevSSB02,SafronovaJD99}.
    
    The final value for the scalar polarizability that we report, $\alpha = -268.82 \: (30) \ a_0^3$ is the sum of all the contributions listed in column nine of the table.  The uncertainty $\delta \alpha = 0.30 \ a_0^3$ is the quadrature sum of the uncertainties listed in the tenth column.  Note that the primary uncertainties now come from the uncertainties of the E1 matrix elements $\langle 6s_{1/2} || r || 8p_{3/2} \rangle $ and $\langle 6s_{1/2} || r || 7p_{3/2} \rangle $, and the tail contributions $\alpha_{n>12}$.
    Our calculated value of $\alpha$ is in agreement with prior calculations of $\alpha$ using the same sum-over-states method~\cite{BlundellSJ92, SafronovaJD99, VasilyevSSB02}, but the 0.11\% precision of the current determination is a significant improvement.

\begin{table}[tb]
    \centering
    \caption{This table lists several determinations of $\beta$ since 1997, and we have bolded the two highest precision determinations. The previous value of $\beta$ with the best precision combines a measurement in 1999 by Bennett et al. of $M1_{hf}/\beta$ and the calculation in 2000 of $M1_{hf}$. The determinations labeled ``Sum over states $(\alpha)$'' combine a calculation of $\alpha$ and the high precision measurement of $\alpha/\beta$ \cite{ChoWBRW1997}. In this table, we have also listed our direct calculation of $\beta$ through a sum-over-states method, which has a large uncertainty due to cancellation of terms.} \label{table:betavalues}
    \begin{tabular}{ll|l|l}
    \multicolumn{1}{c}{Year} & \multicolumn{1}{c}{Authors}   & \multicolumn{1}{c}{Remarks} & \multicolumn{1}{c}{$\beta \ (a_0^3)$} \\
    \hline
    % 2019 &                  & Weighted average				& 27.054 (31) \\
    \textbf{2019} & \textbf{This work}  & \textbf{Sum over states $ \bm{(\alpha)}$} 	& \textbf{27.139 (42)} \\
    %2019 & This work                 & Sum over states $(\beta)$ 	& 27.01 (23) \\
    % 	 & 			 		& 	& \\
    2002 & Dzu02 \cite{DzubaFG02} 		    & Sum over states $(\alpha)$ 	& 27.15 (11) \\
    2002 & Vas02 \cite{VasilyevSSB02}  	& Sum over states $(\alpha)$	& 27.22 (11) \\
    \textbf{2000} & \textbf{Dzu00} \cite{DzubaF00} & \textbf{$\bm{M1_{hf}}$ calculation} & \textbf{26.957 (51)} \\
    1999 & Ben99 \cite{BennettW99}   		& $M1_{hf}/\beta$ expt          & 27.024 (80) \\
    1999 & Saf99 \cite{SafronovaJD99}	& Sum over states $(\alpha)$	& 27.11 (22) \\
    1999 & Saf99 \cite{SafronovaJD99}	& Sum over states $(\beta)$		& 27.16 \\
    1997 & Dzu97 \cite{DzubaFS97}  		    & Sum over states $(\alpha)$ 	& 27.15 (13) \\
    1992 & Blu92 \cite{BlundellSJ92}     & Sum over states $(\beta)$ 	& 27.0 (2) \\    
    \end{tabular}
\end{table}

    \begin{figure}[b!]
\begin{center}
 \includegraphics[width=9cm]{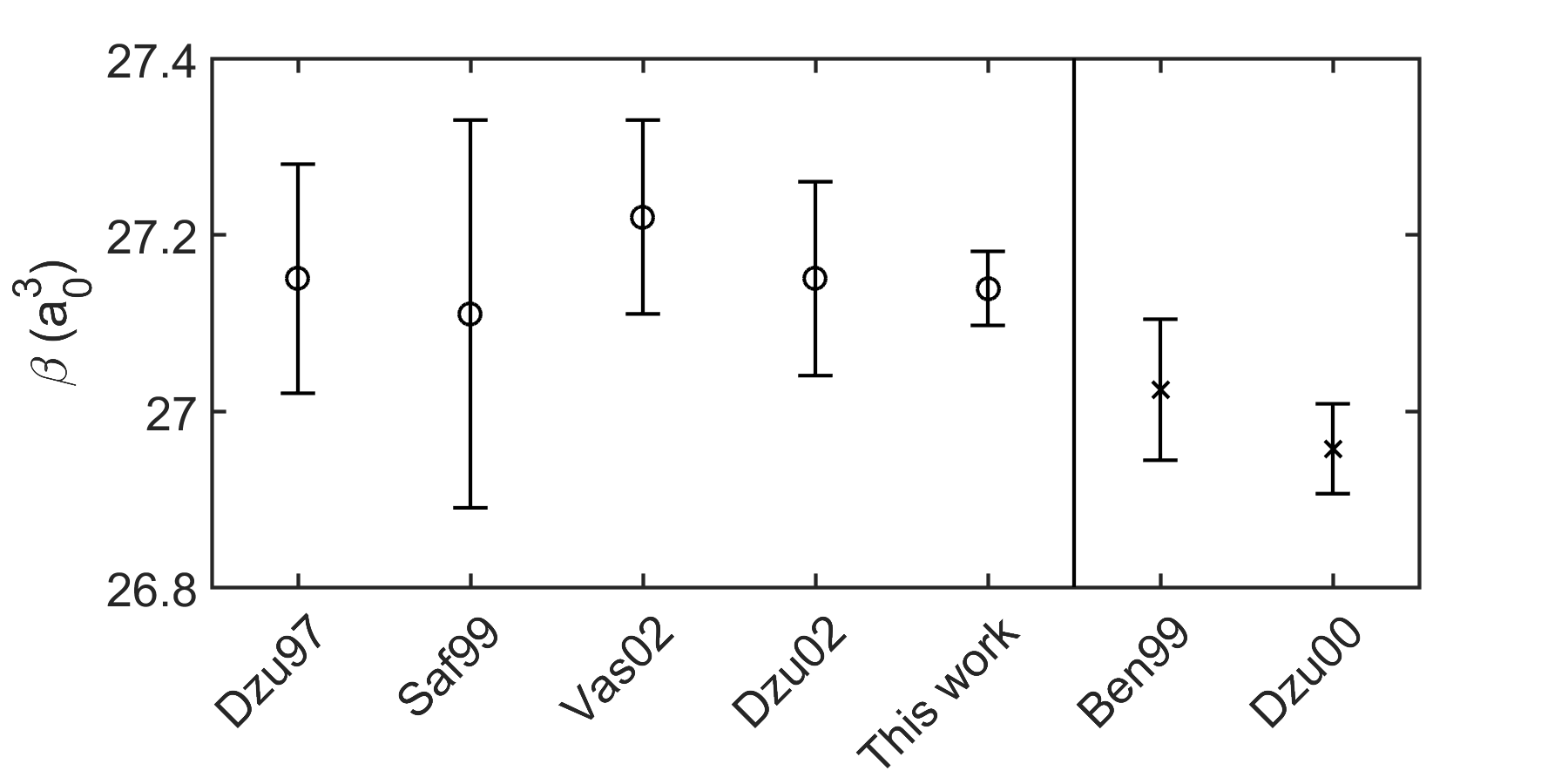} \\
\end{center}
    \caption{A summary of the current status of $\beta$ determinations. The $\beta$ values shown on the left (o) are determined through a sum-over-states of $\alpha$ and the ratio $\alpha/\beta$. The two values on the right (x) are determined through an experimental determination of $M1_{hf}/\beta$ and a theory value of $M1_{hf}$. Refer to Table~\ref{table:betavalues} for references to these values.}
	  \label{fig:BetaPlot}
\end{figure}
From $\alpha$, we use the measured value of $\alpha / \beta = 9.905 \ (11)$ \cite{ChoWBRW1997} to derive \begin{equation}\label{eq:betaresult}
        \beta = 27.139 \ (42) \ a_0^3.
    \end{equation}
We list this result, along with previous determinations of $\beta$ in Table \ref{table:betavalues}, and show these data graphically in Fig.~\ref{fig:BetaPlot}.
The previous best determination of $\beta$, shown in bold font in Table \ref{table:betavalues}, comes from a calculation of the hyperfine changing contribution to the magnetic dipole matrix element $M1_{hf} = 0.8074(8) \times 10^{-5}~\mu_B/c$ \cite{DzubaF00}, thought to be accurate to $0.1\%$, and the measurement of $M1_{hf} / \beta = -5.6195(91)$ V/cm \cite{BennettW99}. This results in $\beta = 26.957 \ (51) \ a_0^3$. These two results differ from one another by $0.182 \ a_0^3 ~(0.67 \%)$, which is larger than the sum of their uncertainties $0.093 \ a_0^3~(0.34 \%)$.  The uncertainty in the new value is slightly smaller than that of the previous best value.  We have also calculated  $\beta = 27.01 \ (23) \ a_0^3$
directly from the E1 data displayed in Table \ref{table:alphasumoverstates} using the sum-over-states expression in Eq.~(40) of Ref.~\cite{BlundellSJ92}.  This value is in agreement with Eq.~(\ref{eq:betaresult}), but with much larger uncertainty due to significant cancellations between terms.
    
The new determination of the vector polarizability has an important implication for $Im ( \mathcal{E}_{\rm PNC})$.  The best measurement of $Im ( \mathcal{E}_{\rm PNC})$ to date is the measurement in 1997 of 
    % $Im$E_{PNC} / \beta = 1.5935 (56)$~mV/cm
    \begin{equation}
    \frac{Im ( \mathcal{E}_{\rm PNC})}{\beta} = 1.5935 (56)~\mathrm{mV/cm}
    \end{equation}
by Wood {\it et al.}~\cite{WoodBCMRTW97}.  (In the following, we base our analysis solely on this value, rather than the 2005 measurement of $Im ( \mathcal{E}_{\rm PNC}) / \beta = 1.538 (40)$ mV/cm by Guena {\it et al.}~\cite{GuenaLB05}.)

To extract the weak charge $Q_w$ of the cesium nucleus from a measurement of $Im ( \mathcal{E}_{\rm PNC})$, we need theoretical calculations of the proportionality $k_{PNC}$ between $Im ( \mathcal{E}_{\rm PNC})$ and $Q_w$.
Many-body calculations done by \cite{PorsevBD09,PorsevBD10} determine 
    \begin{equation}\label{eq:kPorsevBD09}
    Im ( \mathcal{E}_{\rm PNC}) = 0.8906 (24) \times 10^{-11} |e|a_0 \left( -Q_w / N \right).     
    \end{equation}
The authors use the coupled-cluster method with full single, double and valence triple excitations considered. They also accounted for Breit, quantum electrodynamics (QED), and neutron skin corrections. The claimed $0.27 \%$ uncertainty was obtained by comparison of calculations of energies, electric dipole amplitudes and hyperfine constants. Using Eq.~(\ref{eq:kPorsevBD09}) and our value of $\beta$ results in 
    \begin{equation} \label{eq:qw1}
    Q_w = - 73.66 (28)_e (20)_t \ ,
    \end{equation}
    where the experimental (e) and theoretical (t) uncertainties are indicated separately.  This value of the weak charge is $\sim 1.2~\sigma$ larger than the standard model value~\cite{RPP2018} 
    \begin{equation} \label{eq:qwsm}
    Q_{SM}^{2018} = - 73.23(1).
    \end{equation}
    
Dzuba {\it et al.}~\cite{DzubaBFR12,RobertsDF2013} introduced corrections to the core and tail contributions to $Im ( \mathcal{E}_{\rm PNC})$ in Refs.~\cite{PorsevBD09,PorsevBD10} and determined 
    \begin{equation}\label{eq:kDzubaBRF12}
    Im ( \mathcal{E}_{\rm PNC}) = 0.8977 (40) \times 10^{-11} |e|a_0 \left( -Q_w / N \right),
    \end{equation}
in disagreement with Eq.~(\ref{eq:kPorsevBD09}), but in excellent agreement with their earlier results~\cite{DzubaFG02, FlambaumG05}.
Combining Eq.~(\ref{eq:kDzubaBRF12}) with our value of $\beta$ results in the value of 
    \begin{equation} \label{eq:qw2}
    Q_w = - 73.07 (28)_e (33)_t,
    \end{equation}
    $\sim 0.3~\sigma$ less than $Q^{2018}_{SM}$. 
    
\begin{figure}
\begin{center}
 \includegraphics[width=9cm]{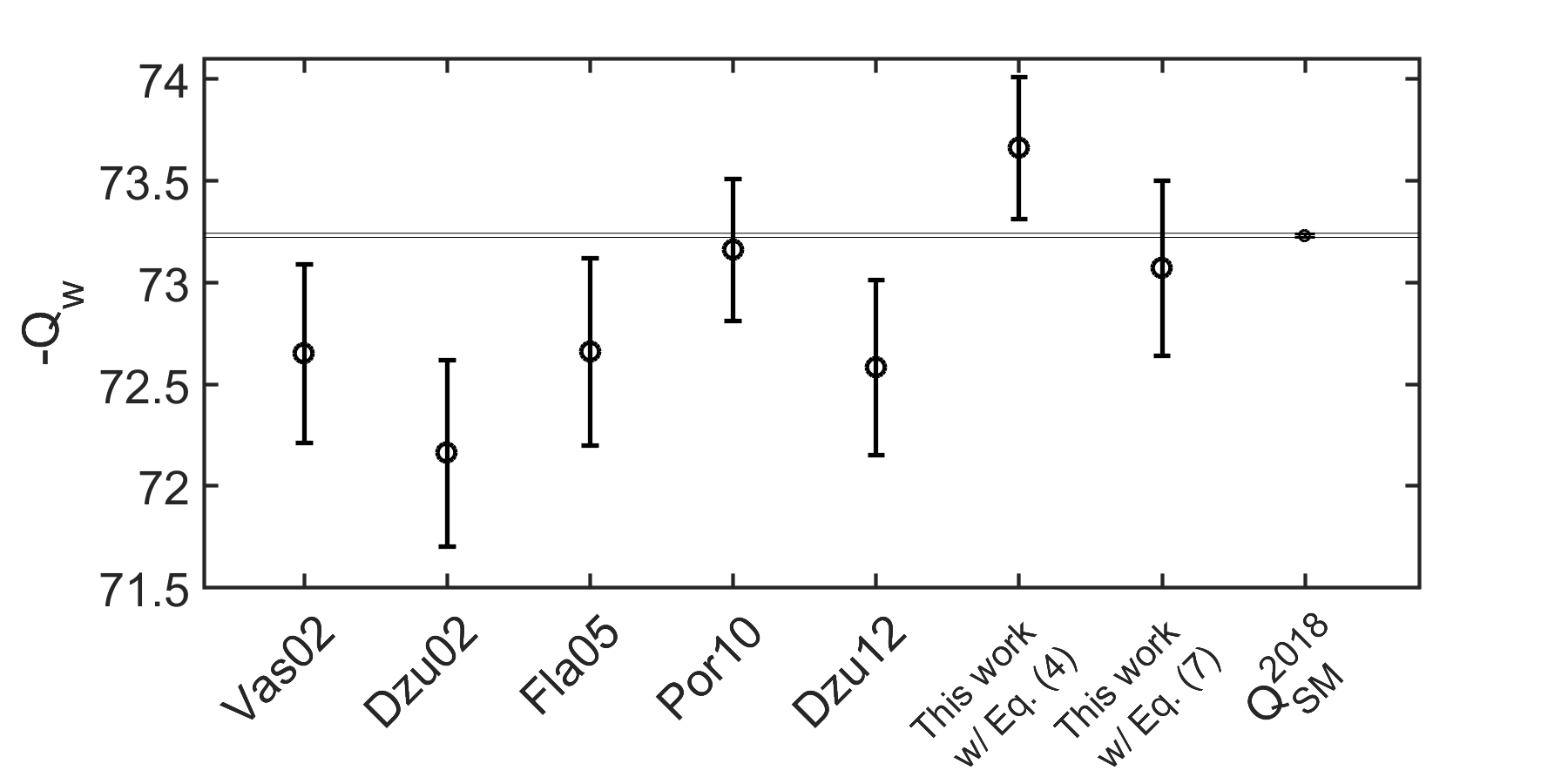} \\
\end{center}
    \caption{A summary of the current status of $Q_w$ determinations. 
    The two horizontal lines denote the Standard Model prediction $Q^{2018}_{SM}$ \cite{RPP2018}. Past determinations are Vas02 \cite{VasilyevSSB02}, Dzu02 \cite{DzubaFG02}, Fla05 \cite{FlambaumG05}, Por10 \cite{PorsevBD10}, Dzu12 \cite{DzubaBFR12}.}
	  \label{fig:QwPlot}
\end{figure}

We show in Fig.~\ref{fig:QwPlot} the various determinations of $Q_w$ since 2002~\cite{VasilyevSSB02, DzubaFG02, FlambaumG05, PorsevBD09, PorsevBD10, DzubaBFR12}. 
The datapoint labeled $Q_{SM}^{2018}$ and the two horizontal lines denote the Standard Model prediction and its uncertainty~\cite{RPP2018}.  We note plans to resolve the differences between Eqs.~(\ref{eq:kPorsevBD09}) and (\ref{eq:kDzubaBRF12}) through a unified calculation of all contributions (principal, tail, and core) to $Im(\mathcal{E}_{PNC})$~\cite{WiemanD19}. 

In conclusion, we report a new, high-precision determination of the scalar ($\alpha$) and vector ($\beta$) polarizabilities of the cesium $6s \rightarrow 7s$ transition. This was achieved using precise values of E1 matrix elements between the lowest energy levels of cesium, which we determined from a combination of measurements and calculations. From that, we report new values for the weak charge of the cesium nucleus $Q_w$. There are still unresolved differences between the two most recent values of the vector polarizability $\beta$, which call for new calculations and/or measurements to address this issue.  We note that any further improvement to the determination of $\alpha$ will require high precision measurements of a few key E1 matrix elements identified above, or alternatively, a direct laboratory determination of $\alpha$.  Furthermore, any improvement to the value of $\beta$ as determined through the method described here will require a new laboratory measurement of $\alpha/\beta$, since the uncertainty of the current value of this ratio is of magnitude comparable to that of $\alpha$.

This material is based upon work supported by the National Science Foundation under Grant Number PHY-1460899.

\bibliography{biblio}

 \newcommand{\noop}[1]{}
\begin{thebibliography}{58}%
\makeatletter
\providecommand \@ifxundefined [1]{%
 \@ifx{#1\undefined}
}%
\providecommand \@ifnum [1]{%
 \ifnum #1\expandafter \@firstoftwo
 \else \expandafter \@secondoftwo
 \fi
}%
\providecommand \@ifx [1]{%
 \ifx #1\expandafter \@firstoftwo
 \else \expandafter \@secondoftwo
 \fi
}%
\providecommand \natexlab [1]{#1}%
\providecommand \enquote  [1]{``#1''}%
\providecommand \bibnamefont  [1]{#1}%
\providecommand \bibfnamefont [1]{#1}%
\providecommand \citenamefont [1]{#1}%
\providecommand \href@noop [0]{\@secondoftwo}%
\providecommand \href [0]{\begingroup \@sanitize@url \@href}%
\providecommand \@href[1]{\@@startlink{#1}\@@href}%
\providecommand \@@href[1]{\endgroup#1\@@endlink}%
\providecommand \@sanitize@url [0]{\catcode `\\12\catcode `\$12\catcode
  `\&12\catcode `\#12\catcode `\^12\catcode `\_12\catcode `\%12\relax}%
\providecommand \@@startlink[1]{}%
\providecommand \@@endlink[0]{}%
\providecommand \url  [0]{\begingroup\@sanitize@url \@url }%
\providecommand \@url [1]{\endgroup\@href {#1}{\urlprefix }}%
\providecommand \urlprefix  [0]{URL }%
\providecommand \Eprint [0]{\href }%
\providecommand \doibase [0]{http://dx.doi.org/}%
\providecommand \selectlanguage [0]{\@gobble}%
\providecommand \bibinfo  [0]{\@secondoftwo}%
\providecommand \bibfield  [0]{\@secondoftwo}%
\providecommand \translation [1]{[#1]}%
\providecommand \BibitemOpen [0]{}%
\providecommand \bibitemStop [0]{}%
\providecommand \bibitemNoStop [0]{.\EOS\space}%
\providecommand \EOS [0]{\spacefactor3000\relax}%
\providecommand \BibitemShut  [1]{\csname bibitem#1\endcsname}%
\let\auto@bib@innerbib\@empty
%</preamble>
\bibitem [{\citenamefont {{Bouchiat, M. A.}}\ and\ \citenamefont {{Bouchiat,
  C.}}(1974)}]{BouchiatB74a}%
  \BibitemOpen
  \bibfield  {author} {\bibinfo {author} {\bibnamefont {{Bouchiat, M. A.}}}\
  and\ \bibinfo {author} {\bibnamefont {{Bouchiat, C.}}},\ }\href {\doibase
  10.1051/jphys:019740035012089900} {\bibfield  {journal} {\bibinfo  {journal}
  {J. Phys. France}\ }\textbf {\bibinfo {volume} {35}},\ \bibinfo {pages} {899}
  (\bibinfo {year} {1974})}\BibitemShut {NoStop}%
\bibitem [{\citenamefont {{Bouchiat, M. A.}}\ and\ \citenamefont {{Bouchiat,
  C.}}(1975)}]{BouchiatB74b}%
  \BibitemOpen
  \bibfield  {author} {\bibinfo {author} {\bibnamefont {{Bouchiat, M. A.}}}\
  and\ \bibinfo {author} {\bibnamefont {{Bouchiat, C.}}},\ }\href {\doibase
  10.1051/jphys:01975003606049300} {\bibfield  {journal} {\bibinfo  {journal}
  {J. Phys. France}\ }\textbf {\bibinfo {volume} {36}},\ \bibinfo {pages} {493}
  (\bibinfo {year} {1975})}\BibitemShut {NoStop}%
\bibitem [{\citenamefont {Porsev}\ \emph {et~al.}(2009)\citenamefont {Porsev},
  \citenamefont {Beloy},\ and\ \citenamefont {Derevianko}}]{PorsevBD09}%
  \BibitemOpen
  \bibfield  {author} {\bibinfo {author} {\bibfnamefont {S.~G.}\ \bibnamefont
  {Porsev}}, \bibinfo {author} {\bibfnamefont {K.}~\bibnamefont {Beloy}}, \
  and\ \bibinfo {author} {\bibfnamefont {A.}~\bibnamefont {Derevianko}},\
  }\href {\doibase 10.1103/PhysRevLett.102.181601} {\bibfield  {journal}
  {\bibinfo  {journal} {Phys. Rev. Lett.}\ }\textbf {\bibinfo {volume} {102}},\
  \bibinfo {pages} {181601} (\bibinfo {year} {2009})}\BibitemShut {NoStop}%
\bibitem [{\citenamefont {Porsev}\ \emph {et~al.}(2010)\citenamefont {Porsev},
  \citenamefont {Beloy},\ and\ \citenamefont {Derevianko}}]{PorsevBD10}%
  \BibitemOpen
  \bibfield  {author} {\bibinfo {author} {\bibfnamefont {S.~G.}\ \bibnamefont
  {Porsev}}, \bibinfo {author} {\bibfnamefont {K.}~\bibnamefont {Beloy}}, \
  and\ \bibinfo {author} {\bibfnamefont {A.}~\bibnamefont {Derevianko}},\
  }\href {\doibase 10.1103/PhysRevD.82.036008} {\bibfield  {journal} {\bibinfo
  {journal} {Phys. Rev. D}\ }\textbf {\bibinfo {volume} {82}},\ \bibinfo
  {pages} {036008} (\bibinfo {year} {2010})}\BibitemShut {NoStop}%
\bibitem [{\citenamefont {Dzuba}\ \emph {et~al.}(2012)\citenamefont {Dzuba},
  \citenamefont {Berengut}, \citenamefont {Flambaum},\ and\ \citenamefont
  {Roberts}}]{DzubaBFR12}%
  \BibitemOpen
  \bibfield  {author} {\bibinfo {author} {\bibfnamefont {V.~A.}\ \bibnamefont
  {Dzuba}}, \bibinfo {author} {\bibfnamefont {J.~C.}\ \bibnamefont {Berengut}},
  \bibinfo {author} {\bibfnamefont {V.~V.}\ \bibnamefont {Flambaum}}, \ and\
  \bibinfo {author} {\bibfnamefont {B.}~\bibnamefont {Roberts}},\ }\href
  {\doibase 10.1103/PhysRevLett.109.203003} {\bibfield  {journal} {\bibinfo
  {journal} {Phys. Rev. Lett.}\ }\textbf {\bibinfo {volume} {109}},\ \bibinfo
  {pages} {203003} (\bibinfo {year} {2012})}\BibitemShut {NoStop}%
\bibitem [{\citenamefont {Erler}\ and\ \citenamefont
  {Langacker}(2000)}]{ErlerL00}%
  \BibitemOpen
  \bibfield  {author} {\bibinfo {author} {\bibfnamefont {J.}~\bibnamefont
  {Erler}}\ and\ \bibinfo {author} {\bibfnamefont {P.}~\bibnamefont
  {Langacker}},\ }\href {\\
  https://link.aps.org/doi/10.1103/PhysRevLett.84.212} {\bibfield  {journal}
  {\bibinfo  {journal} {Physical Review Letters}\ }\textbf {\bibinfo {volume}
  {84}},\ \bibinfo {pages} {212} (\bibinfo {year} {2000})}\BibitemShut
  {NoStop}%
\bibitem [{\citenamefont {Diener}\ \emph {et~al.}(2012)\citenamefont {Diener},
  \citenamefont {Godfrey},\ and\ \citenamefont {Turan}}]{DienerGT12}%
  \BibitemOpen
  \bibfield  {author} {\bibinfo {author} {\bibfnamefont {R.}~\bibnamefont
  {Diener}}, \bibinfo {author} {\bibfnamefont {S.}~\bibnamefont {Godfrey}}, \
  and\ \bibinfo {author} {\bibfnamefont {I.}~\bibnamefont {Turan}},\ }\href
  {\doibase 10.1103/PhysRevD.86.115017} {\bibfield  {journal} {\bibinfo
  {journal} {Phys. Rev. D}\ }\textbf {\bibinfo {volume} {86}},\ \bibinfo
  {pages} {115017} (\bibinfo {year} {2012})}\BibitemShut {NoStop}%
\bibitem [{\citenamefont {Boehm}\ and\ \citenamefont {Fayet}(2004)}]{BoehmF04}%
  \BibitemOpen
  \bibfield  {author} {\bibinfo {author} {\bibfnamefont {C.}~\bibnamefont
  {Boehm}}\ and\ \bibinfo {author} {\bibfnamefont {P.}~\bibnamefont {Fayet}},\
  }\href {https://doi.org/10.1016/j.nuclphysb.2004.01.015} {\bibfield
  {journal} {\bibinfo  {journal} {Nuclear Physics B}\ }\textbf {\bibinfo
  {volume} {683}},\ \bibinfo {pages} {219} (\bibinfo {year}
  {2004})}\BibitemShut {NoStop}%
\bibitem [{\citenamefont {Bouchiat}\ and\ \citenamefont
  {Fayet}(2005)}]{BouchiatF05}%
  \BibitemOpen
  \bibfield  {author} {\bibinfo {author} {\bibfnamefont {C.}~\bibnamefont
  {Bouchiat}}\ and\ \bibinfo {author} {\bibfnamefont {P.}~\bibnamefont
  {Fayet}},\ }\href {https://doi.org/10.1016/j.physletb.2004.12.065} {\bibfield
   {journal} {\bibinfo  {journal} {Physics Letters B}\ }\textbf {\bibinfo
  {volume} {608}},\ \bibinfo {pages} {87} (\bibinfo {year} {2005})}\BibitemShut
  {NoStop}%
\bibitem [{\citenamefont {Dzuba}\ \emph {et~al.}(2017)\citenamefont {Dzuba},
  \citenamefont {Flambaum},\ and\ \citenamefont {Stadnik}}]{DzubaFS17}%
  \BibitemOpen
  \bibfield  {author} {\bibinfo {author} {\bibfnamefont {V.~A.}\ \bibnamefont
  {Dzuba}}, \bibinfo {author} {\bibfnamefont {V.~V.}\ \bibnamefont {Flambaum}},
  \ and\ \bibinfo {author} {\bibfnamefont {Y.~V.}\ \bibnamefont {Stadnik}},\
  }\href {\doibase 10.1103/PhysRevLett.119.223201} {\bibfield  {journal}
  {\bibinfo  {journal} {Phys. Rev. Lett.}\ }\textbf {\bibinfo {volume} {119}},\
  \bibinfo {pages} {223201} (\bibinfo {year} {2017})}\BibitemShut {NoStop}%
\bibitem [{\citenamefont {Stadnik}\ \emph {et~al.}(2018)\citenamefont
  {Stadnik}, \citenamefont {Dzuba},\ and\ \citenamefont
  {Flambaum}}]{StadnikDF18}%
  \BibitemOpen
  \bibfield  {author} {\bibinfo {author} {\bibfnamefont {Y.~V.}\ \bibnamefont
  {Stadnik}}, \bibinfo {author} {\bibfnamefont {V.~A.}\ \bibnamefont {Dzuba}},
  \ and\ \bibinfo {author} {\bibfnamefont {V.~V.}\ \bibnamefont {Flambaum}},\
  }\href {\doibase 10.1103/PhysRevLett.120.013202} {\bibfield  {journal}
  {\bibinfo  {journal} {Phys. Rev. Lett.}\ }\textbf {\bibinfo {volume} {120}},\
  \bibinfo {pages} {013202} (\bibinfo {year} {2018})}\BibitemShut {NoStop}%
\bibitem [{\citenamefont {Roberts}\ \emph
  {et~al.}(2014{\natexlab{a}})\citenamefont {Roberts}, \citenamefont {Stadnik},
  \citenamefont {Dzuba}, \citenamefont {Flambaum}, \citenamefont {Leefer},\
  and\ \citenamefont {Budker}}]{RobertsSDFLB14b}%
  \BibitemOpen
  \bibfield  {author} {\bibinfo {author} {\bibfnamefont {B.~M.}\ \bibnamefont
  {Roberts}}, \bibinfo {author} {\bibfnamefont {Y.~V.}\ \bibnamefont
  {Stadnik}}, \bibinfo {author} {\bibfnamefont {V.~A.}\ \bibnamefont {Dzuba}},
  \bibinfo {author} {\bibfnamefont {V.~V.}\ \bibnamefont {Flambaum}}, \bibinfo
  {author} {\bibfnamefont {N.}~\bibnamefont {Leefer}}, \ and\ \bibinfo {author}
  {\bibfnamefont {D.}~\bibnamefont {Budker}},\ }\href {\doibase
  10.1103/PhysRevD.90.096005} {\bibfield  {journal} {\bibinfo  {journal} {Phys.
  Rev. D}\ }\textbf {\bibinfo {volume} {90}},\ \bibinfo {pages} {096005}
  (\bibinfo {year} {2014}{\natexlab{a}})}\BibitemShut {NoStop}%
\bibitem [{\citenamefont {Roberts}\ \emph
  {et~al.}(2014{\natexlab{b}})\citenamefont {Roberts}, \citenamefont {Stadnik},
  \citenamefont {Dzuba}, \citenamefont {Flambaum}, \citenamefont {Leefer},\
  and\ \citenamefont {Budker}}]{RobertsSDFLB14a}%
  \BibitemOpen
  \bibfield  {author} {\bibinfo {author} {\bibfnamefont {B.~M.}\ \bibnamefont
  {Roberts}}, \bibinfo {author} {\bibfnamefont {Y.~V.}\ \bibnamefont
  {Stadnik}}, \bibinfo {author} {\bibfnamefont {V.~A.}\ \bibnamefont {Dzuba}},
  \bibinfo {author} {\bibfnamefont {V.~V.}\ \bibnamefont {Flambaum}}, \bibinfo
  {author} {\bibfnamefont {N.}~\bibnamefont {Leefer}}, \ and\ \bibinfo {author}
  {\bibfnamefont {D.}~\bibnamefont {Budker}},\ }\href {\doibase
  10.1103/PhysRevLett.113.081601} {\bibfield  {journal} {\bibinfo  {journal}
  {Phys. Rev. Lett.}\ }\textbf {\bibinfo {volume} {113}},\ \bibinfo {pages}
  {081601} (\bibinfo {year} {2014}{\natexlab{b}})}\BibitemShut {NoStop}%
\bibitem [{\citenamefont {Stadnik}\ and\ \citenamefont
  {Flambaum}(2014{\natexlab{a}})}]{StadnikF14a}%
  \BibitemOpen
  \bibfield  {author} {\bibinfo {author} {\bibfnamefont {Y.~V.}\ \bibnamefont
  {Stadnik}}\ and\ \bibinfo {author} {\bibfnamefont {V.~V.}\ \bibnamefont
  {Flambaum}},\ }\href {\doibase 10.1103/PhysRevD.89.043522} {\bibfield
  {journal} {\bibinfo  {journal} {Phys. Rev. D}\ }\textbf {\bibinfo {volume}
  {89}},\ \bibinfo {pages} {043522} (\bibinfo {year}
  {2014}{\natexlab{a}})}\BibitemShut {NoStop}%
\bibitem [{\citenamefont {Stadnik}\ and\ \citenamefont
  {Flambaum}(2014{\natexlab{b}})}]{StadnikF14b}%
  \BibitemOpen
  \bibfield  {author} {\bibinfo {author} {\bibfnamefont {Y.~V.}\ \bibnamefont
  {Stadnik}}\ and\ \bibinfo {author} {\bibfnamefont {V.~V.}\ \bibnamefont
  {Flambaum}},\ }\href {https://doi.org/10.1142/S0217732314400070} {\bibfield
  {journal} {\bibinfo  {journal} {Modern Physics Letters A}\ }\textbf {\bibinfo
  {volume} {29}},\ \bibinfo {pages} {1440007} (\bibinfo {year}
  {2014}{\natexlab{b}})}\BibitemShut {NoStop}%
\bibitem [{\citenamefont {Davoudiasl}\ \emph {et~al.}(2012)\citenamefont
  {Davoudiasl}, \citenamefont {Lee},\ and\ \citenamefont
  {Marciano}}]{DavoudiaslLM12}%
  \BibitemOpen
  \bibfield  {author} {\bibinfo {author} {\bibfnamefont {H.}~\bibnamefont
  {Davoudiasl}}, \bibinfo {author} {\bibfnamefont {H.-S.}\ \bibnamefont {Lee}},
  \ and\ \bibinfo {author} {\bibfnamefont {W.~J.}\ \bibnamefont {Marciano}},\
  }\href {\doibase 10.1103/PhysRevLett.109.031802} {\bibfield  {journal}
  {\bibinfo  {journal} {Phys. Rev. Lett.}\ }\textbf {\bibinfo {volume} {109}},\
  \bibinfo {pages} {031802} (\bibinfo {year} {2012})}\BibitemShut {NoStop}%
\bibitem [{\citenamefont {Davoudiasl}\ and\ \citenamefont
  {Lewis}(2014)}]{DavoudiaslL14}%
  \BibitemOpen
  \bibfield  {author} {\bibinfo {author} {\bibfnamefont {H.}~\bibnamefont
  {Davoudiasl}}\ and\ \bibinfo {author} {\bibfnamefont {I.~M.}\ \bibnamefont
  {Lewis}},\ }\href {\doibase 10.1103/PhysRevD.89.055026} {\bibfield  {journal}
  {\bibinfo  {journal} {Phys. Rev. D}\ }\textbf {\bibinfo {volume} {89}},\
  \bibinfo {pages} {055026} (\bibinfo {year} {2014})}\BibitemShut {NoStop}%
\bibitem [{\citenamefont {Davoudiasl}\ \emph {et~al.}(2014)\citenamefont
  {Davoudiasl}, \citenamefont {Lee},\ and\ \citenamefont
  {Marciano}}]{DavoudiaslLM14}%
  \BibitemOpen
  \bibfield  {author} {\bibinfo {author} {\bibfnamefont {H.}~\bibnamefont
  {Davoudiasl}}, \bibinfo {author} {\bibfnamefont {H.-S.}\ \bibnamefont {Lee}},
  \ and\ \bibinfo {author} {\bibfnamefont {W.~J.}\ \bibnamefont {Marciano}},\
  }\href {\doibase 10.1103/PhysRevD.89.095006} {\bibfield  {journal} {\bibinfo
  {journal} {Phys. Rev. D}\ }\textbf {\bibinfo {volume} {89}},\ \bibinfo
  {pages} {095006} (\bibinfo {year} {2014})}\BibitemShut {NoStop}%
\bibitem [{\citenamefont {Wood}\ \emph {et~al.}(1997)\citenamefont {Wood},
  \citenamefont {Bennett}, \citenamefont {Cho}, \citenamefont {Masterson},
  \citenamefont {Roberts}, \citenamefont {Tanner},\ and\ \citenamefont
  {Wieman}}]{WoodBCMRTW97}%
  \BibitemOpen
  \bibfield  {author} {\bibinfo {author} {\bibfnamefont {C.~S.}\ \bibnamefont
  {Wood}}, \bibinfo {author} {\bibfnamefont {S.~C.}\ \bibnamefont {Bennett}},
  \bibinfo {author} {\bibfnamefont {D.}~\bibnamefont {Cho}}, \bibinfo {author}
  {\bibfnamefont {B.~P.}\ \bibnamefont {Masterson}}, \bibinfo {author}
  {\bibfnamefont {J.~L.}\ \bibnamefont {Roberts}}, \bibinfo {author}
  {\bibfnamefont {C.~E.}\ \bibnamefont {Tanner}}, \ and\ \bibinfo {author}
  {\bibfnamefont {C.~E.}\ \bibnamefont {Wieman}},\ }\href {\doibase
  10.1126/science.275.5307.1759} {\bibfield  {journal} {\bibinfo  {journal}
  {Science}\ }\textbf {\bibinfo {volume} {275}},\ \bibinfo {pages} {1759}
  (\bibinfo {year} {1997})}\BibitemShut {NoStop}%
\bibitem [{\citenamefont {Dzuba}\ \emph {et~al.}(1989)\citenamefont {Dzuba},
  \citenamefont {Flambaum},\ and\ \citenamefont {Sushkov}}]{DzubaFS89}%
  \BibitemOpen
  \bibfield  {author} {\bibinfo {author} {\bibfnamefont {V.}~\bibnamefont
  {Dzuba}}, \bibinfo {author} {\bibfnamefont {V.}~\bibnamefont {Flambaum}}, \
  and\ \bibinfo {author} {\bibfnamefont {O.}~\bibnamefont {Sushkov}},\ }\href
  {\doibase https://doi.org/10.1016/0375-9601(89)90777-9} {\bibfield  {journal}
  {\bibinfo  {journal} {Physics Letters A}\ }\textbf {\bibinfo {volume}
  {141}},\ \bibinfo {pages} {147 } (\bibinfo {year} {1989})}\BibitemShut
  {NoStop}%
\bibitem [{\citenamefont {Blundell}\ \emph {et~al.}(1991)\citenamefont
  {Blundell}, \citenamefont {Johnson},\ and\ \citenamefont
  {Sapirstein}}]{BlundellJS91}%
  \BibitemOpen
  \bibfield  {author} {\bibinfo {author} {\bibfnamefont {S.~A.}\ \bibnamefont
  {Blundell}}, \bibinfo {author} {\bibfnamefont {W.~R.}\ \bibnamefont
  {Johnson}}, \ and\ \bibinfo {author} {\bibfnamefont {J.}~\bibnamefont
  {Sapirstein}},\ }\href {\doibase 10.1103/PhysRevA.43.3407} {\bibfield
  {journal} {\bibinfo  {journal} {Phys. Rev. A}\ }\textbf {\bibinfo {volume}
  {43}},\ \bibinfo {pages} {3407} (\bibinfo {year} {1991})}\BibitemShut
  {NoStop}%
\bibitem [{\citenamefont {Blundell}\ \emph {et~al.}(1992)\citenamefont
  {Blundell}, \citenamefont {Sapirstein},\ and\ \citenamefont
  {Johnson}}]{BlundellSJ92}%
  \BibitemOpen
  \bibfield  {author} {\bibinfo {author} {\bibfnamefont {S.~A.}\ \bibnamefont
  {Blundell}}, \bibinfo {author} {\bibfnamefont {J.}~\bibnamefont
  {Sapirstein}}, \ and\ \bibinfo {author} {\bibfnamefont {W.~R.}\ \bibnamefont
  {Johnson}},\ }\href {\doibase 10.1103/PhysRevD.45.1602} {\bibfield  {journal}
  {\bibinfo  {journal} {Phys. Rev. D}\ }\textbf {\bibinfo {volume} {45}},\
  \bibinfo {pages} {1602} (\bibinfo {year} {1992})}\BibitemShut {NoStop}%
\bibitem [{\citenamefont {Derevianko}(2000)}]{Derevianko00}%
  \BibitemOpen
  \bibfield  {author} {\bibinfo {author} {\bibfnamefont {A.}~\bibnamefont
  {Derevianko}},\ }\href {\doibase 10.1103/PhysRevLett.85.1618} {\bibfield
  {journal} {\bibinfo  {journal} {Phys. Rev. Lett.}\ }\textbf {\bibinfo
  {volume} {85}},\ \bibinfo {pages} {1618} (\bibinfo {year}
  {2000})}\BibitemShut {NoStop}%
\bibitem [{\citenamefont {Dzuba}\ \emph {et~al.}(2001)\citenamefont {Dzuba},
  \citenamefont {Flambaum},\ and\ \citenamefont {Ginges}}]{DzubaFG01}%
  \BibitemOpen
  \bibfield  {author} {\bibinfo {author} {\bibfnamefont {V.~A.}\ \bibnamefont
  {Dzuba}}, \bibinfo {author} {\bibfnamefont {V.~V.}\ \bibnamefont {Flambaum}},
  \ and\ \bibinfo {author} {\bibfnamefont {J.~S.~M.}\ \bibnamefont {Ginges}},\
  }\href {\doibase 10.1103/PhysRevA.63.062101} {\bibfield  {journal} {\bibinfo
  {journal} {Phys. Rev. A}\ }\textbf {\bibinfo {volume} {63}},\ \bibinfo
  {pages} {062101} (\bibinfo {year} {2001})}\BibitemShut {NoStop}%
\bibitem [{\citenamefont {Johnson}\ \emph {et~al.}(2001)\citenamefont
  {Johnson}, \citenamefont {Bednyakov},\ and\ \citenamefont
  {Soff}}]{JohnsonBS01}%
  \BibitemOpen
  \bibfield  {author} {\bibinfo {author} {\bibfnamefont {W.~R.}\ \bibnamefont
  {Johnson}}, \bibinfo {author} {\bibfnamefont {I.}~\bibnamefont {Bednyakov}},
  \ and\ \bibinfo {author} {\bibfnamefont {G.}~\bibnamefont {Soff}},\ }\href
  {\doibase 10.1103/PhysRevLett.87.233001} {\bibfield  {journal} {\bibinfo
  {journal} {Phys. Rev. Lett.}\ }\textbf {\bibinfo {volume} {87}},\ \bibinfo
  {pages} {233001} (\bibinfo {year} {2001})}\BibitemShut {NoStop}%
\bibitem [{\citenamefont {Kozlov}\ \emph {et~al.}(2001)\citenamefont {Kozlov},
  \citenamefont {Porsev},\ and\ \citenamefont {Tupitsyn}}]{KozlovPT01}%
  \BibitemOpen
  \bibfield  {author} {\bibinfo {author} {\bibfnamefont {M.~G.}\ \bibnamefont
  {Kozlov}}, \bibinfo {author} {\bibfnamefont {S.~G.}\ \bibnamefont {Porsev}},
  \ and\ \bibinfo {author} {\bibfnamefont {I.~I.}\ \bibnamefont {Tupitsyn}},\
  }\href {\doibase 10.1103/PhysRevLett.86.3260} {\bibfield  {journal} {\bibinfo
   {journal} {Phys. Rev. Lett.}\ }\textbf {\bibinfo {volume} {86}},\ \bibinfo
  {pages} {3260} (\bibinfo {year} {2001})}\BibitemShut {NoStop}%
\bibitem [{\citenamefont {Dzuba}\ \emph {et~al.}(2002)\citenamefont {Dzuba},
  \citenamefont {Flambaum},\ and\ \citenamefont {Ginges}}]{DzubaFG02}%
  \BibitemOpen
  \bibfield  {author} {\bibinfo {author} {\bibfnamefont {V.~A.}\ \bibnamefont
  {Dzuba}}, \bibinfo {author} {\bibfnamefont {V.~V.}\ \bibnamefont {Flambaum}},
  \ and\ \bibinfo {author} {\bibfnamefont {J.~S.~M.}\ \bibnamefont {Ginges}},\
  }\href {\doibase 10.1103/PhysRevD.66.076013} {\bibfield  {journal} {\bibinfo
  {journal} {Phys. Rev. D}\ }\textbf {\bibinfo {volume} {66}},\ \bibinfo
  {pages} {076013} (\bibinfo {year} {2002})}\BibitemShut {NoStop}%
\bibitem [{\citenamefont {Flambaum}\ and\ \citenamefont
  {Ginges}(2005)}]{FlambaumG05}%
  \BibitemOpen
  \bibfield  {author} {\bibinfo {author} {\bibfnamefont {V.~V.}\ \bibnamefont
  {Flambaum}}\ and\ \bibinfo {author} {\bibfnamefont {J.~S.~M.}\ \bibnamefont
  {Ginges}},\ }\href {\doibase 10.1103/PhysRevA.72.052115} {\bibfield
  {journal} {\bibinfo  {journal} {Phys. Rev. A}\ }\textbf {\bibinfo {volume}
  {72}},\ \bibinfo {pages} {052115} (\bibinfo {year} {2005})}\BibitemShut
  {NoStop}%
\bibitem [{\citenamefont {Roberts}\ \emph {et~al.}(2013)\citenamefont
  {Roberts}, \citenamefont {Dzuba},\ and\ \citenamefont
  {Flambaum}}]{RobertsDF2013}%
  \BibitemOpen
  \bibfield  {author} {\bibinfo {author} {\bibfnamefont {B.~M.}\ \bibnamefont
  {Roberts}}, \bibinfo {author} {\bibfnamefont {V.~A.}\ \bibnamefont {Dzuba}},
  \ and\ \bibinfo {author} {\bibfnamefont {V.~V.}\ \bibnamefont {Flambaum}},\
  }\href {\doibase 10.1103/PhysRevA.87.054502} {\bibfield  {journal} {\bibinfo
  {journal} {Phys. Rev. A}\ }\textbf {\bibinfo {volume} {87}},\ \bibinfo
  {pages} {054502} (\bibinfo {year} {2013})}\BibitemShut {NoStop}%
\bibitem [{\citenamefont {Wieman}\ and\ \citenamefont
  {Derevianko}(2019)}]{WiemanD19}%
  \BibitemOpen
  \bibfield  {author} {\bibinfo {author} {\bibfnamefont {C.}~\bibnamefont
  {Wieman}}\ and\ \bibinfo {author} {\bibfnamefont {A.}~\bibnamefont
  {Derevianko}},\ }\href@noop {} {\  (\bibinfo {year} {2019})},\ \Eprint
  {http://arxiv.org/abs/1904.00281} {arXiv:1904.00281 [physics.atom-ph]}
  \BibitemShut {NoStop}%
%%CITATION = ARXIV:1904.00281;%%
\bibitem [{\citenamefont {Dzuba}\ and\ \citenamefont
  {Flambaum}(2000)}]{DzubaF00}%
  \BibitemOpen
  \bibfield  {author} {\bibinfo {author} {\bibfnamefont {V.~A.}\ \bibnamefont
  {Dzuba}}\ and\ \bibinfo {author} {\bibfnamefont {V.~V.}\ \bibnamefont
  {Flambaum}},\ }\href {\doibase 10.1103/PhysRevA.62.052101} {\bibfield
  {journal} {\bibinfo  {journal} {Phys. Rev. A}\ }\textbf {\bibinfo {volume}
  {62}},\ \bibinfo {pages} {052101} (\bibinfo {year} {2000})}\BibitemShut
  {NoStop}%
\bibitem [{\citenamefont {Bennett}\ and\ \citenamefont
  {Wieman}(1999)}]{BennettW99}%
  \BibitemOpen
  \bibfield  {author} {\bibinfo {author} {\bibfnamefont {S.~C.}\ \bibnamefont
  {Bennett}}\ and\ \bibinfo {author} {\bibfnamefont {C.~E.}\ \bibnamefont
  {Wieman}},\ }\href {\doibase 10.1103/PhysRevLett.82.2484} {\bibfield
  {journal} {\bibinfo  {journal} {Physical Review Letters}\ }\textbf {\bibinfo
  {volume} {82}},\ \bibinfo {pages} {2484} (\bibinfo {year}
  {1999})}\BibitemShut {NoStop}%
\bibitem [{\citenamefont {Safronova}\ \emph {et~al.}(1999)\citenamefont
  {Safronova}, \citenamefont {Johnson},\ and\ \citenamefont
  {Derevianko}}]{SafronovaJD99}%
  \BibitemOpen
  \bibfield  {author} {\bibinfo {author} {\bibfnamefont {M.~S.}\ \bibnamefont
  {Safronova}}, \bibinfo {author} {\bibfnamefont {W.~R.}\ \bibnamefont
  {Johnson}}, \ and\ \bibinfo {author} {\bibfnamefont {A.}~\bibnamefont
  {Derevianko}},\ }\href {\doibase 10.1103/PhysRevA.60.4476} {\bibfield
  {journal} {\bibinfo  {journal} {Phys. Rev. A}\ }\textbf {\bibinfo {volume}
  {60}},\ \bibinfo {pages} {4476} (\bibinfo {year} {1999})}\BibitemShut
  {NoStop}%
\bibitem [{\citenamefont {Vasilyev}\ \emph {et~al.}(2002)\citenamefont
  {Vasilyev}, \citenamefont {Savukov}, \citenamefont {Safronova},\ and\
  \citenamefont {Berry}}]{VasilyevSSB02}%
  \BibitemOpen
  \bibfield  {author} {\bibinfo {author} {\bibfnamefont {A.~A.}\ \bibnamefont
  {Vasilyev}}, \bibinfo {author} {\bibfnamefont {I.~M.}\ \bibnamefont
  {Savukov}}, \bibinfo {author} {\bibfnamefont {M.~S.}\ \bibnamefont
  {Safronova}}, \ and\ \bibinfo {author} {\bibfnamefont {H.~G.}\ \bibnamefont
  {Berry}},\ }\href {\doibase 10.1103/PhysRevA.66.020101} {\bibfield  {journal}
  {\bibinfo  {journal} {Phys. Rev. A}\ }\textbf {\bibinfo {volume} {66}},\
  \bibinfo {pages} {020101(R)} (\bibinfo {year} {2002})}\BibitemShut {NoStop}%
\bibitem [{\citenamefont {Cho}\ \emph {et~al.}(1997)\citenamefont {Cho},
  \citenamefont {Wood}, \citenamefont {Bennett}, \citenamefont {Roberts},\ and\
  \citenamefont {Wieman}}]{ChoWBRW1997}%
  \BibitemOpen
  \bibfield  {author} {\bibinfo {author} {\bibfnamefont {D.}~\bibnamefont
  {Cho}}, \bibinfo {author} {\bibfnamefont {C.~S.}\ \bibnamefont {Wood}},
  \bibinfo {author} {\bibfnamefont {S.~C.}\ \bibnamefont {Bennett}}, \bibinfo
  {author} {\bibfnamefont {J.~L.}\ \bibnamefont {Roberts}}, \ and\ \bibinfo
  {author} {\bibfnamefont {C.~E.}\ \bibnamefont {Wieman}},\ }\href {\doibase
  10.1103/PhysRevA.55.1007} {\bibfield  {journal} {\bibinfo  {journal} {Phys.
  Rev. A}\ }\textbf {\bibinfo {volume} {55}},\ \bibinfo {pages} {1007}
  (\bibinfo {year} {1997})}\BibitemShut {NoStop}%
\bibitem [{\citenamefont {{Bouchiat, M.A.}}\ \emph {et~al.}(1984)\citenamefont
  {{Bouchiat, M.A.}}, \citenamefont {{Guena, J.}},\ and\ \citenamefont
  {{Pottier, L.}}}]{BouchiatGP84}%
  \BibitemOpen
  \bibfield  {author} {\bibinfo {author} {\bibnamefont {{Bouchiat, M.A.}}},
  \bibinfo {author} {\bibnamefont {{Guena, J.}}}, \ and\ \bibinfo {author}
  {\bibnamefont {{Pottier, L.}}},\ }\href {\doibase
  10.1051/jphyslet:019840045011052300} {\bibfield  {journal} {\bibinfo
  {journal} {J. Physique Lett.}\ }\textbf {\bibinfo {volume} {45}},\ \bibinfo
  {pages} {523} (\bibinfo {year} {1984})}\BibitemShut {NoStop}%
\bibitem [{\citenamefont {Tanner}\ \emph {et~al.}(1992)\citenamefont {Tanner},
  \citenamefont {Livingston}, \citenamefont {Rafac}, \citenamefont {Serpa},
  \citenamefont {Kukla}, \citenamefont {Berry}, \citenamefont {Young},\ and\
  \citenamefont {Kurtz}}]{TannerLRSKBYK92}%
  \BibitemOpen
  \bibfield  {author} {\bibinfo {author} {\bibfnamefont {C.~E.}\ \bibnamefont
  {Tanner}}, \bibinfo {author} {\bibfnamefont {A.~E.}\ \bibnamefont
  {Livingston}}, \bibinfo {author} {\bibfnamefont {R.~J.}\ \bibnamefont
  {Rafac}}, \bibinfo {author} {\bibfnamefont {F.~G.}\ \bibnamefont {Serpa}},
  \bibinfo {author} {\bibfnamefont {K.~W.}\ \bibnamefont {Kukla}}, \bibinfo
  {author} {\bibfnamefont {H.~G.}\ \bibnamefont {Berry}}, \bibinfo {author}
  {\bibfnamefont {L.}~\bibnamefont {Young}}, \ and\ \bibinfo {author}
  {\bibfnamefont {C.~A.}\ \bibnamefont {Kurtz}},\ }\href {\doibase
  10.1103/PhysRevLett.69.2765} {\bibfield  {journal} {\bibinfo  {journal}
  {Phys. Rev. Lett.}\ }\textbf {\bibinfo {volume} {69}},\ \bibinfo {pages}
  {2765} (\bibinfo {year} {1992})}\BibitemShut {NoStop}%
\bibitem [{\citenamefont {Young}\ \emph {et~al.}(1994)\citenamefont {Young},
  \citenamefont {Hill}, \citenamefont {Sibener}, \citenamefont {Price},
  \citenamefont {Tanner}, \citenamefont {Wieman},\ and\ \citenamefont
  {Leone}}]{YoungHSPTWL94}%
  \BibitemOpen
  \bibfield  {author} {\bibinfo {author} {\bibfnamefont {L.}~\bibnamefont
  {Young}}, \bibinfo {author} {\bibfnamefont {W.~T.}\ \bibnamefont {Hill}},
  \bibinfo {author} {\bibfnamefont {S.~J.}\ \bibnamefont {Sibener}}, \bibinfo
  {author} {\bibfnamefont {S.~D.}\ \bibnamefont {Price}}, \bibinfo {author}
  {\bibfnamefont {C.~E.}\ \bibnamefont {Tanner}}, \bibinfo {author}
  {\bibfnamefont {C.~E.}\ \bibnamefont {Wieman}}, \ and\ \bibinfo {author}
  {\bibfnamefont {S.~R.}\ \bibnamefont {Leone}},\ }\href {\doibase
  10.1103/PhysRevA.50.2174} {\bibfield  {journal} {\bibinfo  {journal} {Phys.
  Rev. A}\ }\textbf {\bibinfo {volume} {50}},\ \bibinfo {pages} {2174}
  (\bibinfo {year} {1994})}\BibitemShut {NoStop}%
\bibitem [{\citenamefont {Rafac}\ and\ \citenamefont
  {Tanner}(1998)}]{RafacT98}%
  \BibitemOpen
  \bibfield  {author} {\bibinfo {author} {\bibfnamefont {R.~J.}\ \bibnamefont
  {Rafac}}\ and\ \bibinfo {author} {\bibfnamefont {C.~E.}\ \bibnamefont
  {Tanner}},\ }\href {\doibase 10.1103/PhysRevA.58.1087} {\bibfield  {journal}
  {\bibinfo  {journal} {Phys. Rev. A}\ }\textbf {\bibinfo {volume} {58}},\
  \bibinfo {pages} {1087} (\bibinfo {year} {1998})}\BibitemShut {NoStop}%
\bibitem [{\citenamefont {Rafac}\ \emph {et~al.}(1999)\citenamefont {Rafac},
  \citenamefont {Tanner}, \citenamefont {Livingston},\ and\ \citenamefont
  {Berry}}]{RafacTLB99}%
  \BibitemOpen
  \bibfield  {author} {\bibinfo {author} {\bibfnamefont {R.~J.}\ \bibnamefont
  {Rafac}}, \bibinfo {author} {\bibfnamefont {C.~E.}\ \bibnamefont {Tanner}},
  \bibinfo {author} {\bibfnamefont {A.~E.}\ \bibnamefont {Livingston}}, \ and\
  \bibinfo {author} {\bibfnamefont {H.~G.}\ \bibnamefont {Berry}},\ }\href
  {\doibase 10.1103/PhysRevA.60.3648} {\bibfield  {journal} {\bibinfo
  {journal} {Phys. Rev. A}\ }\textbf {\bibinfo {volume} {60}},\ \bibinfo
  {pages} {3648} (\bibinfo {year} {1999})}\BibitemShut {NoStop}%
\bibitem [{\citenamefont {Bennett}\ \emph {et~al.}(1999)\citenamefont
  {Bennett}, \citenamefont {Roberts},\ and\ \citenamefont
  {Wieman}}]{BennettRW99}%
  \BibitemOpen
  \bibfield  {author} {\bibinfo {author} {\bibfnamefont {S.~C.}\ \bibnamefont
  {Bennett}}, \bibinfo {author} {\bibfnamefont {J.~L.}\ \bibnamefont
  {Roberts}}, \ and\ \bibinfo {author} {\bibfnamefont {C.~E.}\ \bibnamefont
  {Wieman}},\ }\href {\doibase 10.1103/PhysRevA.59.R16} {\bibfield  {journal}
  {\bibinfo  {journal} {Phys. Rev. A}\ }\textbf {\bibinfo {volume} {59}},\
  \bibinfo {pages} {R16} (\bibinfo {year} {1999})}\BibitemShut {NoStop}%
\bibitem [{\citenamefont {Derevianko}\ and\ \citenamefont
  {Porsev}(2002)}]{DereviankoP02a}%
  \BibitemOpen
  \bibfield  {author} {\bibinfo {author} {\bibfnamefont {A.}~\bibnamefont
  {Derevianko}}\ and\ \bibinfo {author} {\bibfnamefont {S.~G.}\ \bibnamefont
  {Porsev}},\ }\href {\doibase 10.1103/PhysRevA.65.053403} {\bibfield
  {journal} {\bibinfo  {journal} {Phys. Rev. A}\ }\textbf {\bibinfo {volume}
  {65}},\ \bibinfo {pages} {053403} (\bibinfo {year} {2002})}\BibitemShut
  {NoStop}%
\bibitem [{\citenamefont {Amini}\ and\ \citenamefont {Gould}(2003)}]{AminiG03}%
  \BibitemOpen
  \bibfield  {author} {\bibinfo {author} {\bibfnamefont {J.~M.}\ \bibnamefont
  {Amini}}\ and\ \bibinfo {author} {\bibfnamefont {H.}~\bibnamefont {Gould}},\
  }\href {\doibase 10.1103/PhysRevLett.91.153001} {\bibfield  {journal}
  {\bibinfo  {journal} {Phys. Rev. Lett.}\ }\textbf {\bibinfo {volume} {91}},\
  \bibinfo {pages} {153001} (\bibinfo {year} {2003})}\BibitemShut {NoStop}%
\bibitem [{\citenamefont {Bouloufa}\ \emph {et~al.}(2007)\citenamefont
  {Bouloufa}, \citenamefont {Crubellier},\ and\ \citenamefont
  {Dulieu}}]{BouloufaCD07}%
  \BibitemOpen
  \bibfield  {author} {\bibinfo {author} {\bibfnamefont {N.}~\bibnamefont
  {Bouloufa}}, \bibinfo {author} {\bibfnamefont {A.}~\bibnamefont
  {Crubellier}}, \ and\ \bibinfo {author} {\bibfnamefont {O.}~\bibnamefont
  {Dulieu}},\ }\href {\doibase 10.1103/PhysRevA.75.052501} {\bibfield
  {journal} {\bibinfo  {journal} {Phys. Rev. A}\ }\textbf {\bibinfo {volume}
  {75}},\ \bibinfo {pages} {052501} (\bibinfo {year} {2007})}\BibitemShut
  {NoStop}%
\bibitem [{\citenamefont {Sell}\ \emph {et~al.}(2011)\citenamefont {Sell},
  \citenamefont {Patterson}, \citenamefont {Ehrenreich}, \citenamefont
  {Brooke}, \citenamefont {Scoville},\ and\ \citenamefont
  {Knize}}]{SellPEBSK11}%
  \BibitemOpen
  \bibfield  {author} {\bibinfo {author} {\bibfnamefont {J.~F.}\ \bibnamefont
  {Sell}}, \bibinfo {author} {\bibfnamefont {B.~M.}\ \bibnamefont {Patterson}},
  \bibinfo {author} {\bibfnamefont {T.}~\bibnamefont {Ehrenreich}}, \bibinfo
  {author} {\bibfnamefont {G.}~\bibnamefont {Brooke}}, \bibinfo {author}
  {\bibfnamefont {J.}~\bibnamefont {Scoville}}, \ and\ \bibinfo {author}
  {\bibfnamefont {R.~J.}\ \bibnamefont {Knize}},\ }\href {\doibase
  10.1103/PhysRevA.84.010501} {\bibfield  {journal} {\bibinfo  {journal} {Phys.
  Rev. A}\ }\textbf {\bibinfo {volume} {84}},\ \bibinfo {pages} {010501(R)}
  (\bibinfo {year} {2011})}\BibitemShut {NoStop}%
\bibitem [{\citenamefont {Zhang}\ \emph {et~al.}(2013)\citenamefont {Zhang},
  \citenamefont {Ma}, \citenamefont {Wu}, \citenamefont {Wang}, \citenamefont
  {Xiao},\ and\ \citenamefont {Jia}}]{ZhangMWWXJ13}%
  \BibitemOpen
  \bibfield  {author} {\bibinfo {author} {\bibfnamefont {Y.}~\bibnamefont
  {Zhang}}, \bibinfo {author} {\bibfnamefont {J.}~\bibnamefont {Ma}}, \bibinfo
  {author} {\bibfnamefont {J.}~\bibnamefont {Wu}}, \bibinfo {author}
  {\bibfnamefont {L.}~\bibnamefont {Wang}}, \bibinfo {author} {\bibfnamefont
  {L.}~\bibnamefont {Xiao}}, \ and\ \bibinfo {author} {\bibfnamefont
  {S.}~\bibnamefont {Jia}},\ }\href {\doibase 10.1103/PhysRevA.87.030503}
  {\bibfield  {journal} {\bibinfo  {journal} {Phys. Rev. A}\ }\textbf {\bibinfo
  {volume} {87}},\ \bibinfo {pages} {030503(R)} (\bibinfo {year}
  {2013})}\BibitemShut {NoStop}%
\bibitem [{\citenamefont {Antypas}\ and\ \citenamefont
  {Elliott}(2013)}]{antypas7p2013}%
  \BibitemOpen
  \bibfield  {author} {\bibinfo {author} {\bibfnamefont {D.}~\bibnamefont
  {Antypas}}\ and\ \bibinfo {author} {\bibfnamefont {D.~S.}\ \bibnamefont
  {Elliott}},\ }\href {\doibase 10.1103/PhysRevA.88.052516} {\bibfield
  {journal} {\bibinfo  {journal} {Phys. Rev. A}\ }\textbf {\bibinfo {volume}
  {88}},\ \bibinfo {pages} {052516} (\bibinfo {year} {2013})}\BibitemShut
  {NoStop}%
\bibitem [{\citenamefont {Borv\'{a}k}(2014)}]{Borvak14}%
  \BibitemOpen
  \bibfield  {author} {\bibinfo {author} {\bibfnamefont {L.}~\bibnamefont
  {Borv\'{a}k}},\ }\emph {\bibinfo {title} {Direct laser absorption
  spectroscopy measurements of transition strengths in cesium}},\ \href@noop {}
  {Ph.D. thesis},\ \bibinfo  {school} {University of Notre Dame} (\bibinfo
  {year} {2014})\BibitemShut {NoStop}%
\bibitem [{\citenamefont {Patterson}\ \emph {et~al.}(2015)\citenamefont
  {Patterson}, \citenamefont {Sell}, \citenamefont {Ehrenreich}, \citenamefont
  {Gearba}, \citenamefont {Brooke}, \citenamefont {Scoville},\ and\
  \citenamefont {Knize}}]{PattersonSEGBSK15}%
  \BibitemOpen
  \bibfield  {author} {\bibinfo {author} {\bibfnamefont {B.~M.}\ \bibnamefont
  {Patterson}}, \bibinfo {author} {\bibfnamefont {J.~F.}\ \bibnamefont {Sell}},
  \bibinfo {author} {\bibfnamefont {T.}~\bibnamefont {Ehrenreich}}, \bibinfo
  {author} {\bibfnamefont {M.~A.}\ \bibnamefont {Gearba}}, \bibinfo {author}
  {\bibfnamefont {G.~M.}\ \bibnamefont {Brooke}}, \bibinfo {author}
  {\bibfnamefont {J.}~\bibnamefont {Scoville}}, \ and\ \bibinfo {author}
  {\bibfnamefont {R.~J.}\ \bibnamefont {Knize}},\ }\href {\doibase
  10.1103/PhysRevA.91.012506} {\bibfield  {journal} {\bibinfo  {journal} {Phys.
  Rev. A}\ }\textbf {\bibinfo {volume} {91}},\ \bibinfo {pages} {012506}
  (\bibinfo {year} {2015})}\BibitemShut {NoStop}%
\bibitem [{\citenamefont {Gregoire}\ \emph {et~al.}(2015)\citenamefont
  {Gregoire}, \citenamefont {Hromada}, \citenamefont {Holmgren}, \citenamefont
  {Trubko},\ and\ \citenamefont {Cronin}}]{GregoireHHTC15}%
  \BibitemOpen
  \bibfield  {author} {\bibinfo {author} {\bibfnamefont {M.~D.}\ \bibnamefont
  {Gregoire}}, \bibinfo {author} {\bibfnamefont {I.}~\bibnamefont {Hromada}},
  \bibinfo {author} {\bibfnamefont {W.~F.}\ \bibnamefont {Holmgren}}, \bibinfo
  {author} {\bibfnamefont {R.}~\bibnamefont {Trubko}}, \ and\ \bibinfo {author}
  {\bibfnamefont {A.~D.}\ \bibnamefont {Cronin}},\ }\href {\doibase
  10.1103/PhysRevA.92.052513} {\bibfield  {journal} {\bibinfo  {journal} {Phys.
  Rev. A}\ }\textbf {\bibinfo {volume} {92}},\ \bibinfo {pages} {052513}
  (\bibinfo {year} {2015})}\BibitemShut {NoStop}%
\bibitem [{\citenamefont {Toh}\ \emph {et~al.}(2018)\citenamefont {Toh},
  \citenamefont {Jaramillo-Villegas}, \citenamefont {Glotzbach}, \citenamefont
  {Quirk}, \citenamefont {Stevenson}, \citenamefont {Choi}, \citenamefont
  {Weiner},\ and\ \citenamefont {Elliott}}]{TohJGQSCWE18}%
  \BibitemOpen
  \bibfield  {author} {\bibinfo {author} {\bibfnamefont {G.}~\bibnamefont
  {Toh}}, \bibinfo {author} {\bibfnamefont {J.~A.}\ \bibnamefont
  {Jaramillo-Villegas}}, \bibinfo {author} {\bibfnamefont {N.}~\bibnamefont
  {Glotzbach}}, \bibinfo {author} {\bibfnamefont {J.}~\bibnamefont {Quirk}},
  \bibinfo {author} {\bibfnamefont {I.~C.}\ \bibnamefont {Stevenson}}, \bibinfo
  {author} {\bibfnamefont {J.}~\bibnamefont {Choi}}, \bibinfo {author}
  {\bibfnamefont {A.~M.}\ \bibnamefont {Weiner}}, \ and\ \bibinfo {author}
  {\bibfnamefont {D.~S.}\ \bibnamefont {Elliott}},\ }\href {\doibase
  10.1103/PhysRevA.97.052507} {\bibfield  {journal} {\bibinfo  {journal} {Phys.
  Rev. A}\ }\textbf {\bibinfo {volume} {97}},\ \bibinfo {pages} {052507}
  (\bibinfo {year} {2018})}\BibitemShut {NoStop}%
\bibitem [{\citenamefont {Toh}\ \emph {et~al.}(2019)\citenamefont {Toh},
  \citenamefont {Damitz}, \citenamefont {Glotzbach}, \citenamefont {Quirk},
  \citenamefont {Stevenson}, \citenamefont {Choi}, \citenamefont {Safronova},\
  and\ \citenamefont {Elliott}}]{TohDGQSCSE19}%
  \BibitemOpen
  \bibfield  {author} {\bibinfo {author} {\bibfnamefont {G.}~\bibnamefont
  {Toh}}, \bibinfo {author} {\bibfnamefont {A.}~\bibnamefont {Damitz}},
  \bibinfo {author} {\bibfnamefont {N.}~\bibnamefont {Glotzbach}}, \bibinfo
  {author} {\bibfnamefont {J.}~\bibnamefont {Quirk}}, \bibinfo {author}
  {\bibfnamefont {I.~C.}\ \bibnamefont {Stevenson}}, \bibinfo {author}
  {\bibfnamefont {J.}~\bibnamefont {Choi}}, \bibinfo {author} {\bibfnamefont
  {M.~S.}\ \bibnamefont {Safronova}}, \ and\ \bibinfo {author} {\bibfnamefont
  {D.~S.}\ \bibnamefont {Elliott}},\ }\href {\doibase
  10.1103/PhysRevA.99.032504} {\bibfield  {journal} {\bibinfo  {journal} {Phys.
  Rev. A}\ }\textbf {\bibinfo {volume} {99}},\ \bibinfo {pages} {032504}
  (\bibinfo {year} {2019})}\BibitemShut {NoStop}%
\bibitem [{\citenamefont {Damitz}\ \emph {et~al.}(tted)\citenamefont {Damitz},
  \citenamefont {Toh}, \citenamefont {Putney}, \citenamefont {Tanner},\ and\
  \citenamefont {Elliott}}]{DamitzTPTE18a}%
  \BibitemOpen
  \bibfield  {author} {\bibinfo {author} {\bibfnamefont {A.}~\bibnamefont
  {Damitz}}, \bibinfo {author} {\bibfnamefont {G.}~\bibnamefont {Toh}},
  \bibinfo {author} {\bibfnamefont {E.}~\bibnamefont {Putney}}, \bibinfo
  {author} {\bibfnamefont {C.~E.}\ \bibnamefont {Tanner}}, \ and\ \bibinfo
  {author} {\bibfnamefont {D.~S.}\ \bibnamefont {Elliott}},\ }\href@noop {} {\
  (\bibinfo {year} {\noop{3001}submitted})}\BibitemShut {NoStop}%
\bibitem [{\citenamefont {Safronova}\ \emph {et~al.}(2016)\citenamefont
  {Safronova}, \citenamefont {Safronova},\ and\ \citenamefont
  {Clark}}]{SafronovaSC16}%
  \BibitemOpen
  \bibfield  {author} {\bibinfo {author} {\bibfnamefont {M.~S.}\ \bibnamefont
  {Safronova}}, \bibinfo {author} {\bibfnamefont {U.~I.}\ \bibnamefont
  {Safronova}}, \ and\ \bibinfo {author} {\bibfnamefont {C.~W.}\ \bibnamefont
  {Clark}},\ }\href {\doibase 10.1103/PhysRevA.94.012505} {\bibfield  {journal}
  {\bibinfo  {journal} {Phys. Rev. A}\ }\textbf {\bibinfo {volume} {94}},\
  \bibinfo {pages} {012505} (\bibinfo {year} {2016})}\BibitemShut {NoStop}%
\bibitem [{\citenamefont {Kramida}\ \emph {et~al.}(2019)\citenamefont
  {Kramida}, \citenamefont {Ralchenko}, \citenamefont {Reader},\ and\
  \citenamefont {{NIST ASD Team}}}]{kramida2016nist}%
  \BibitemOpen
  \bibfield  {author} {\bibinfo {author} {\bibfnamefont {A.}~\bibnamefont
  {Kramida}}, \bibinfo {author} {\bibfnamefont {Y.}~\bibnamefont {Ralchenko}},
  \bibinfo {author} {\bibfnamefont {J.}~\bibnamefont {Reader}}, \ and\ \bibinfo
  {author} {\bibnamefont {{NIST ASD Team}}},\ }\href
  {https://physics.nist.gov/asd} {\bibfield  {journal} {\bibinfo  {journal}
  {NIST Atomic Spectra Database (version 5.6.1)}\ } (\bibinfo {year}
  {2019})}\BibitemShut {NoStop}%
\bibitem [{\citenamefont {Dzuba}\ \emph {et~al.}(1997)\citenamefont {Dzuba},
  \citenamefont {Flambaum},\ and\ \citenamefont {Sushkov}}]{DzubaFS97}%
  \BibitemOpen
  \bibfield  {author} {\bibinfo {author} {\bibfnamefont {V.~A.}\ \bibnamefont
  {Dzuba}}, \bibinfo {author} {\bibfnamefont {V.~V.}\ \bibnamefont {Flambaum}},
  \ and\ \bibinfo {author} {\bibfnamefont {O.~P.}\ \bibnamefont {Sushkov}},\
  }\href {\doibase 10.1103/PhysRevA.56.R4357} {\bibfield  {journal} {\bibinfo
  {journal} {Phys. Rev. A}\ }\textbf {\bibinfo {volume} {56}},\ \bibinfo
  {pages} {R4357} (\bibinfo {year} {1997})}\BibitemShut {NoStop}%
\bibitem [{\citenamefont {Gu\'ena}\ \emph {et~al.}(2005)\citenamefont
  {Gu\'ena}, \citenamefont {Lintz},\ and\ \citenamefont
  {Bouchiat}}]{GuenaLB05}%
  \BibitemOpen
  \bibfield  {author} {\bibinfo {author} {\bibfnamefont {J.}~\bibnamefont
  {Gu\'ena}}, \bibinfo {author} {\bibfnamefont {M.}~\bibnamefont {Lintz}}, \
  and\ \bibinfo {author} {\bibfnamefont {M.~A.}\ \bibnamefont {Bouchiat}},\
  }\href {\doibase 10.1103/PhysRevA.71.042108} {\bibfield  {journal} {\bibinfo
  {journal} {Phys. Rev. A}\ }\textbf {\bibinfo {volume} {71}},\ \bibinfo
  {pages} {042108} (\bibinfo {year} {2005})}\BibitemShut {NoStop}%
\bibitem [{\citenamefont {Tanabashi~et al.}(2018)}]{RPP2018}%
  \BibitemOpen
  \bibfield  {author} {\bibinfo {author} {\bibfnamefont {M.}~\bibnamefont
  {Tanabashi~et al.}} (\bibinfo {collaboration} {Particle Data Group}),\ }\href
  {\doibase 10.1103/PhysRevD.98.030001} {\bibfield  {journal} {\bibinfo
  {journal} {Phys. Rev. D}\ }\textbf {\bibinfo {volume} {98}},\ \bibinfo
  {pages} {030001} (\bibinfo {year} {2018})}\BibitemShut {NoStop}%
\end{thebibliography}%

\end{document}